# Collective Plasmonic-Molecular Modes in the Strong Coupling Regime

Adi Salomon[1], Robert J. Gordon[1,2], Yehiam Prior[1], Tamar Seideman[3], and Maxim Sukharev[4]

[1]Chemical Physics Department, Weizmann Institute of Science, 76100 Rehovot, Israel

[2]Department of Chemistry, University of Illinois at Chicago, 845 W Taylor Street, Chicago, IL 60680, USA

[3]Department of Chemistry, Northwestern University, 2145 Sheridan Road, Evanston, IL 60208, USA

[4]Department of Applied Sciences and Mathematics, Arizona State University, 6073 S Backus Mall, Mesa, AZ 85212, USA

**Abstract -** We demonstrate strong coupling between molecular excited states and surface plasmon modes of a slit array in a thin metal film. The coupling manifests itself as an anti-crossing behavior of the two newly formed polaritons. As the coupling strength grows, a new mode emerges, which is attributed to long range molecular interactions mediated by the plasmonic field. The new, molecular-like mode repels the polariton states, and leads to an opening of energy gaps both below and above the asymptotic free molecule energy.

The field of nano-plasmonics opened up new possibilities for the study of light-matter interactions at the nano scale, especially when molecules are placed in the regions where the electromagnetic fields are both enhanced and focused to the nano scale [1] by plasmonic resonances [2-4]. Systems comprised of metallic nanostructured arrays are particularly attractive for interaction with photonic molecular excited states, as the plasmonic resonances may be tuned by geometrical factors to exactly match the molecular frequencies [5]. Indeed, strong Interactions between plasmonic and molecular modes have been observed in recent years for both localized [6-9] and delocalized [10-16] plasmonic modes. The interaction of molecules with surface plasmons is similar to that in microcavities [17-20], where the cavity modes dispersion relations are replaced by the plasmonic ones. In the strong coupling regime, energy exchange between the molecular and plasmonic modes is observed, giving rise to two new 'polariton' eigenmodes. Strong coupling manifests itself as an avoided crossing of the polariton modes as a function of the plasmon frequency, with an observed Rabi splitting (RS) which is a non-negligible fraction of the molecular transition frequency [8, 11, 12, 21, 22]. In several experimental studies, a new spectral feature (peak) was observed near the molecular transition, but its nature and physical origin remain to be understood [8, 12, 23]. In several cases additional spectral features were observed, such as reflection peaks assigned by Geiss et al. [24] to quantum dots which were not coupled to the photonic modes, or additional transmission dips observed by Fujita et al. [25] and explained in terms of uncoupled higher-lying polaritonic states.

Here we study the interaction between plasmonic modes excited within an array of slits in a thin silver film and a layer of molecules deposited in close proximity to the surface. We use a self-consistent model and numerically solve the Maxwell-Liouville-von Neumann equations, where the coupling between the molecular excitations and the metallic surface plasmon polariton (SPP) modes is explicitly taken into account. In addition to providing a detailed analysis of the origin of the avoided crossing between the split resonance components as the array spacing varies, we report on the emergence of a new mode. The spectral properties of the new mode are different from those of both the SPP and the molecular modes that led to its onset. We simulate the SPP-molecular interaction in this strong coupling regime for experimentally realizable parameters and explore its dependence on the dipole strength, the molecular density, and the distance of the molecular layer from the surface. We illustrate the conditions under which such new spectral features are expected, and unravel the signatures of the strong coupling regime in the new context of nanoplasmonics. The strong coupling regime is further characterized by a new asymptotic energy gap between the SPP modes and the molecular spectral position, a phenomenon that is directly related to the existence of the new collective mode. In what follows we first describe the results, then elaborate on our theoretical model, and finally conclude with a detailed discussion of our observations.

Consider an infinite array of slits in a thin, 100 nm, silver film as shown in the inset of Fig. 1. The plasmonic response of this slit array may be experimentally tuned by adjusting the array periodicity [26]. For each periodicity, the transmission spectrum is calculated and the plasmonic peak position is derived. A 10 nm molecular layer with its first electronic excited state lying at 2.62 eV is placed above the silver slit array, with or without a spacer layer (see below). The silver dispersion relation is described by the conventional Drude model [27]. The hybrid system is illuminated from above by a white light continuum, and we calculate the transmission spectra. We restrict our treatment to two dimensions and consider only a transverse-electric (TE) incoming plane wave, with its electric field along the x-direction, with the assumption that characteristic cross-sectional geometrical parameters are independent of the y-coordinate (see Fig.1). To model the molecular subsystem, we employ a self-consistent approach based on numerical integration of the coupled Maxwell-Liouville-von Neumann equations. In order to take into account all possible electric field polarizations in the near-field zone, the molecular layer is modeled as a three-level quantum system with two degenerate excited states representing the $p_x$ and $p_z$ orbitals. We adopt the numerical implementation and methodology discussed in detail in ref. [28]. In brief, in spatial regions occupied by molecules we numerically integrate the relevant Maxwell-Liouville-von Neumann equations and evaluate the macroscopic polarization of the combined molecular-plasmonic ensemble. The electrodynamics of the hybrid system is treated by a finite-difference time-domain algorithm [29]. The resulting system of equations is solved numerically on a local cluster [30]. The absorbance spectrum of the molecules and the transmission spectrum of bare slit array are depicted in Fig. 1.

At first the hybrid system is studied at low molecular density, where the molecular dipole-dipole interaction is small, and the electromagnetic field may be assumed to interact individually with each molecule. A set of transmission spectra of the hybrid system, over the relevant energy range and for different array periodicities, is displayed in Fig. 2a. As the array periodicity is tuned, a clear picture of avoided-crossing is observed, both in Fig. 2a, where the actual spectra are shown, and in Fig. 2b, where the peak position of the upper and lower polaritons, $E_u$, and $E_l$, are plotted as a function of the array wave-vector (2π/periodicity). At low periodicity, the upper polariton (blue curve) is 'plasmonic' in nature (having a broad spectral line). As the periodicity is scanned through the molecular resonance, this polariton changes its character and becomes 'molecular' in nature (narrow line). The lower polariton follows the opposite path. At 410 nm periodicity, the two polaritons anti-cross, with a Rabi splitting of 35 meV, reflecting the coupling strength of the molecular-plasmon interaction for the parameters used in this calculation. The solid curves superimposed on the numerically derived anti-crossing eigenmodes are the result of diagonalizing the (2x2) diabatic Hamiltonian, $H = \begin{pmatrix} E_m & \Delta \\ \Delta & E_{pl}(k) \end{pmatrix}$ for an RS value of $2\Delta = 35$ meV, a molecular eigenvalue $E_m$=2.62 eV, and a plasmon mode dispersion, $E_{pl}(k)$, derived from our simulations in absence of molecules. The dispersion relation for the newly formed polaritons, $E_{u,l}(k)$ is given by [17, 18]

$$E_{u,l}(k) = \frac{1}{2}\left\{\left(E_{pl}(k) + E_m\right) \pm \left[4\Delta^2 + \left(E_{pl}(k) - E_m\right)^2\right]^{1/2}\right\} \qquad (1)$$

This model corresponds to the interaction of isolated molecules with the plasmonic modes. While its main predictions are verified experimentally, the model is not expected to be valid for high molecular densities or for strong intermolecular interactions. Moreover, this simple model is stationary, and therefore it is unable to provide information about the lifetime and the emission spectrum of the coupled system [31].

As the molecular density increases, intermolecular dipole-dipole forces and, in particular, polariton-mediated dipole-dipole coupling grows significantly, and the two-level model presented above becomes invalid. Fig. 3a depicts a series of transmission spectra for different molecular densities, ranging from $10^{23}$ molecules/m$^3$ (intermolecular distance of ~50 nm) to $10^{26}$ molecules/m$^3$ (intermolecular distance of ~5 nm). The inset depicts the RS values at resonance as a function of the molecular density. At low densities, the two-level model fits the data well, and the RS values are proportional to the square root of the molecular density [17, 18]. At higher densities, when the intermolecular distance approaches a few nanometers, a new peak appears in the transmission spectrum near the molecular frequency, and deviations from the square root behavior are observed in the density dependence of the RS value.

A similar role is played by the magnitude of the dipole moment, which likewise controls the intermolecular interaction strength. Fig. 3b shows a series of transmission plots for different dipole moment magnitudes, illustrating that both the RS values and the appearance of the new mode are strongly dependent on the molecular dipole moment. The inset shows a linear dependence of the RS on the dipole moment.

Next, we repeat the type of calculation that led to Fig. 2, but this time for a much higher molecular density. Unlike the previous case, where EM field interacted with individual molecules, here, the closer proximity of the molecules enhances the inter-molecule interactions, thus transforming the response into a collective behavior of the entire molecular ensemble. Fig. 4a depicts a series of spectra where the avoided crossing and the extra peak are clearly visible, and Fig. 4b shows the peak positions extracted from the calculations. In Fig. 4b, the two polaritons repel each other further, leading to an observed RS of 150 meV and the emergence of a new peak. The new mode is only slightly dispersive and stays close to the molecular transition frequency as the SP modes are tuned, revealing a molecular, rather than a plasmonic character. Thus, another new feature is now clearly seen. Furthermore, whereas at small coupling strength (e.g., Fig. 2), both upper and lower polaritons approach asymptotically the molecular transition energy, here the behavior is different. The upper polariton opens up an energy gap and asymptotically tends to approach a value that is higher than the molecular transition energy by 70 meV, a very large fraction of the observed RS value of 150 meV. Likewise, the lower polariton develops an energy gap, albeit with a much smaller value (15 meV). These energy gaps are closely related to the appearance of the new peak and serve as another signature of strong coupling.

To further elucidate the origin of the new observations, we proceed by inserting a spacer layer between the molecules and the silver film with a refractive index of 1.0 and a variable thickness. The purpose is to check the dependence of the intermolecular coupling on the strength of the plasmonic field, which decays exponentially with distance from the surface. Fig. 5 is a 3-dimensional plot depicting transmission spectra for a series of distances, ranging from a few nm, where the plasmonic interac-tion is full, to a distance of 300 nm, where the interaction is negligible because of the exponential decay of the SP modes. Two observations are noted; first, the new mode is intense, not only at close proximity to the metal layer but also at distances of 50 nm and 100 nm. As the spacer thickness increases, however, it gradually merges with the upper polariton. Second, the RS separation between the two polaritons decreases exponentially, as is expected for a decaying plasmonic field. The decay of the collective mode with the decrease of the plasmonic field provides further support to dependence of this mode on plasmon-mediated dipole-dipole coupling.

To provide a clear test of the proposed collective nature of the new mode, we performed calculations in which the plasmon-induced dipole-dipole interactions between the molecules are omitted while no other changes are introduced. Technically this is done by replacing the molecular layer with a single two-level system occupying the same volume as molecular layer in which the local electric field is spatially uniform, thereby eliminating any internal dipole-coupling, while maintaining all other interactions intact. The dipole-dipole interactions, which arise in the full model from the field-induced interactions between individual molecules, are therefore absent. Simulations of the transmission spectra of such hybrid system are shown in Fig. 6. Indeed, when molecular dipole-dipole interactions are excluded, the RS value is smaller and the third mode is not observed. Even at significantly higher molecular density, when the RS value is about 200 meV, the third peak is absent. When dipole-dipole interactions between the molecules are included the third mode (peak) is observed and the RS value becomes noticeably larger as shown in Fig. 6.

The dipole-dipole free space coupling is inversely proportional to the third power of the distance between them; for molecules a few nm apart, this interaction is negligible. By contrast, molecules immersed in the strong polaritonic field are coupled via vastly enhanced dipole-dipole interactions induced by the propagating polaritons. Plasmon-mediated dipole-dipole interactions are qualitatively and quantitatively different from the analogous interactions in free space. Unlike the case of space, where the electrostatic interactions delocalize in full 3D, here the interaction takes place in a space of reduced dimensionality. Dipolar interactions in reduced dimensions have been discussed before, near nanoparticles and in waveguides [32], [33], and were shown to give rise to efficient coupling. Here the dipoles lie in a two-dimensional layer, but the plasmon propagation is not isotropic in this 2D space, and the coupling is therefore expected to be an intermediate case between one and two dimensions. In this restricted lower dimensional space, the inverse distance dependence of the interaction is lower than third power. Thus, for shorter distances (higher densities) the interaction energy is higher than for the three dimensional free space. The full analysis of this coupling will be the subject of a future publication.

In most of our calculations we used a three-level model consisting of a ground state and a doubly degenerate excited state. To eliminate the possibility that the observed effect might be an artifact of the degeneracy of the upper level, we repeated our calculation using a two-level model and obtained identical results.

In conclusion, a self-consistent solution of the coupled Maxwell-Liouville-von Neumann equations provides new insights into the strong coupling regime between molecules and SP modes. We show that at higher molecular densities, a plasmon-mediated inter-molecule coupling mechanism exists, which gives rise to a new collective mode of the molecular system, the origin of which is clarified here for the first time. The collective mode results from long-range, plasmon-mediated molecular interactions and is expected in other systems where molecules interact strongly with propagating plasmons. The mechanism leading to the collective behavior gives rise to the new mode and causes a repulsion of the upper polariton and the appearance of an energy gap.

Although a simple model suffices to explain the mode structure, the observed phenomena are general and are expected to play an important role in many realistic systems exhibiting strong interaction between surface plasmons and molecular ensembles. Our model provides tools for tuning the properties of hybrid materials (molecules interacting with a plasmonic array), for applications in areas such as photochemistry [12] [34] and optical device engineering. Better understanding of the prop-erties of mixed plasmon-molecule states may open up the potential for engineering non-linear optical devices.

We acknowledge useful discussions with Ephraim Shahmoon and Joseph Zyss. T.S. is grateful for support by the US Department of Energy (Grant number DE-SC0001785) and the National Science Foundation's MRSEC program (DMR-1121262) at the Materials Research Center of Northwestern University. Y.P. is grateful to the Israel Science Foundation for partial support of this research.

**FIGURES:**

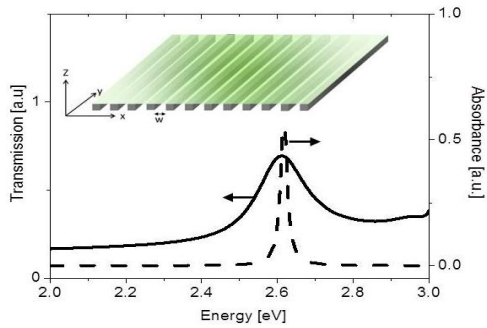

**Figure 1:** The simulated absorption spectrum (dash line) of a 10 nm molecular layer and transmission spectrum (solid line) of bare Ag slit array with the thickness of 100 nm. The molecular transition frequency is 2.62 eV, with transition dipole moment of 25.5 Debye, radiationless lifetime of 1 ps, and molecular density of $3 \times 10^{25}$ m$^{-3}$. The slit array periodicity is 410 nm with slit width of 160 nm. The inset schematically depicts the hybrid system: molecular layer is deposited directly on the Ag slit array.

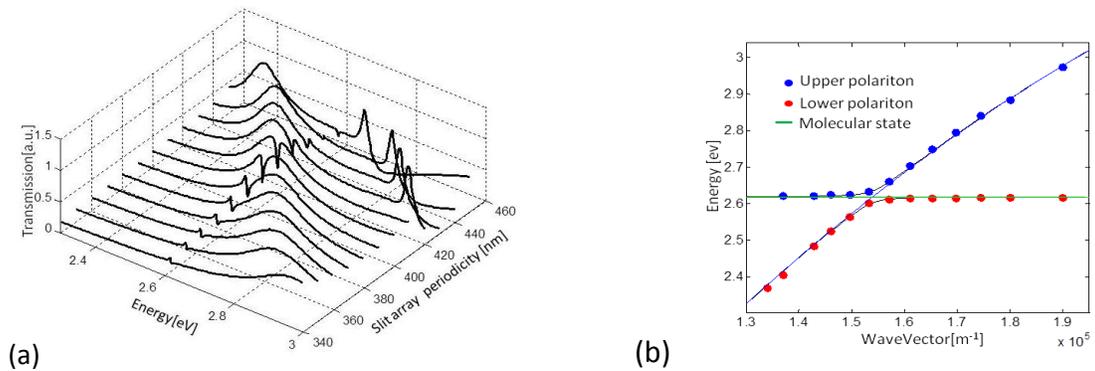

**Figure 2: (a)** Transmission spectra for the series of Ag slit arrays covered by a 10 nm thin film of molecular layer with a density of $10^{24}$ m$^{-3}$. The molecular parameters are as for Fig. 1, but with radiationless lifetime of 300 fs. By changing the slit array periodicity, the SPP modes are tuned to be off and on resonance with respect to the molecular subsystem. **(b)** Positions of the peaks in the panel (a) are plotted as functions of the slit array period. The solid curves superimposed on the numerically derived eigenmodes are the result of diagonalizing the (2x2) Hamiltonian of the system (see eq. (1) in the text).

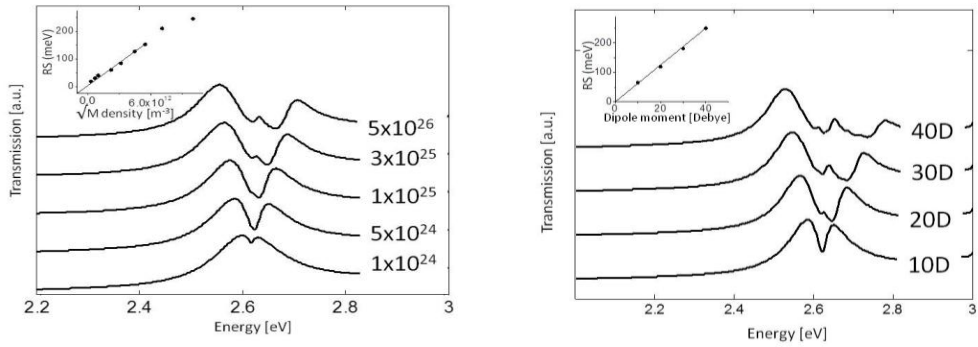

**Figure 3:** Transmission spectra for the Ag slit array with periodicity of 410 nm, covered by a thin film of molecular layer with molecular parameters are as for Fig. 1. When the molecular density is high **(a)** or the transition dipole moment value is large **(b)**, an additional mode (splitting) is observed. Insets: RS values as function of (a) the square root of the molecular densities and (b) as function of the molecular dipole moment values. The curves are displaced for clarity.

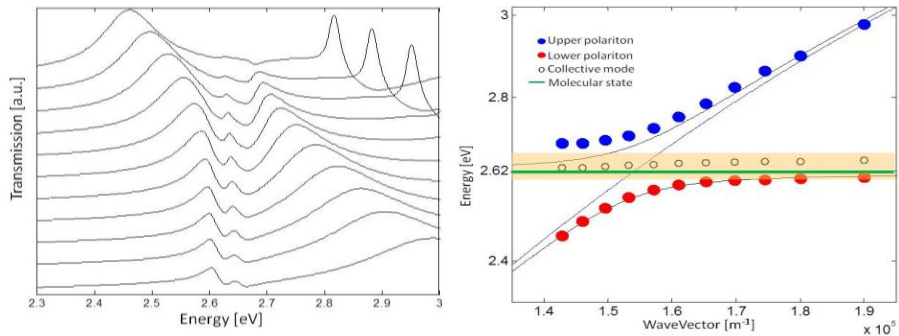

**Figure 4: (a)** Transmission spectra for the series of Ag slit arrays covered by a 10 nm thin film of molecular layer with a density of $3\times10^{25}$ m$^{-3}$. The molecular parameters are as for Fig. 1. An additional mode is clearly seen at about 2.64 eV. **(b)** Anti-crossing behavior of the hybrid system with RS value of 0.15 eV. The peak position of the additional mode barely changes with detuning of the SPP mode. The solid curves superimposed on the numerically derived eigenmodes are the result of diagonalizing the (2x2) Hamiltonian of the system (see eq. (1) in the text). The orange zone indicates an opening of an energy gap between the molecular state and the upper polariton.

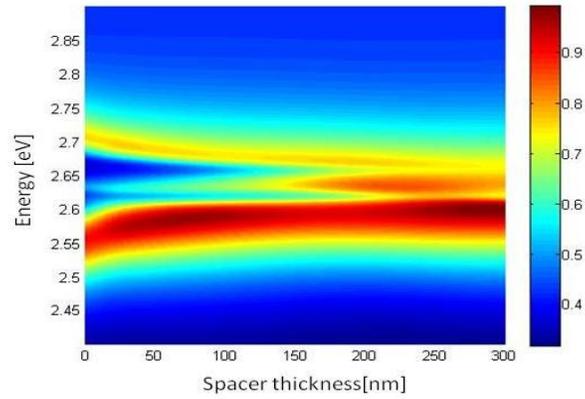

**Figure 5:** Set of transmission spectra for the Ag slit array with slit width of 160 nm and periodicity of 410 nm covered by a spacer with varied thickness of d=0-300 nm and a thin molecular film with the same parameters as for Fig. 1. At d=0-100 nm 3 peaks are observed, whereas at d=300 nm, the three peaks merge and almost no splitting is observed.

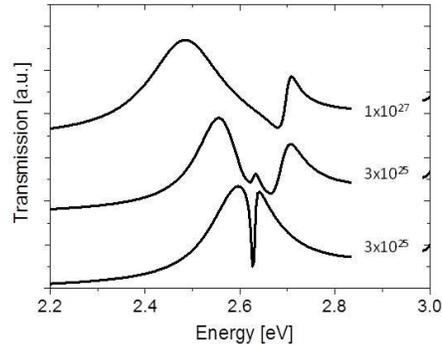

**Figure 6:** simulated transmission spectra for Ag slit array covered by molecules with physical parameters as in Fig. 1. When all interactions are taken into account (middle curve) and at molecular density is $3\times10^{25}$ $m^{-3}$, a third peak close to the molecular transition is observed. When plasmon-induced dipole-dipole interactions are excluded the RS value is smaller and the third peak disappears (bottom). Even at higher molecular density of $10^{27}$ $m^{-3}$, the third peak is not observed.